\documentclass[sort&compress,final,3p,times,twocolumn,fixfloat]{elsarticle}

\journal{Physics Letters B}

\usepackage{graphicx,color}
\usepackage[dvipsnames]{xcolor}
\usepackage[utf8]{inputenc}
\usepackage{txfonts}
\usepackage[colorlinks,citecolor=blue,linktoc=all,linkcolor=cyan]{hyperref}
\usepackage{bm}
\usepackage{amsmath}
\usepackage{amssymb}
\usepackage{booktabs}
\usepackage{float}
\usepackage{xspace} 
\usepackage{mathtools, cuted}
\usepackage[sort&compress, capitalise]{cleveref}
\usepackage{scalerel}
\usepackage[normalem]{ulem}
\usepackage{comment}

\newcommand{\nn}{\nonumber}

\begin{document}
\begin{frontmatter}

\title{GTMDs and the factorization of exclusive double Drell-Yan}

\author[a,b]{Miguel G. Echevarria}
\ead{miguel.garciae@ehu.eus}
\author[c]{Patricia A. Gutierrez Garcia}
\ead{patricgu@ucm.es}
\author[c]{Ignazio Scimemi}
\ead{ignazios@ucm.es}

\address[a]{Department of Physics, University of the Basque Country UPV/EHU,
P.O. Box 644, 48080 Bilbao, Spain}
\address[b]{EHU Quantum Center, University of the Basque Country, UPV/EHU,
P.O. Box 644, 48080 Bilbao, Spain}
\address[c]{Departamento de Física Teórica and IPARCOS, Universidad Complutense de Madrid (UCM),
Plaza Ciencias 1, 28040 Madrid, Spain}

\begin{abstract}
Different exclusive processes have been proposed to access the generalized transverse momentum dependent distributions (GTMDs) with no proof of factorization, which allows to rigorously define the GTMDs.
Using Soft Collinear Effective Theory we derive for the first time the factorization of the differential cross section for the exclusive double Drell-Yan process $\pi N\rightarrow N' \gamma^* \gamma^*\rightarrow N' (\ell^+\ell^-)(\ell^+\ell^-)$, for small transverse momenta of the photons in terms of a perturbatively calculable hard factor, GTMDs and light-cone wave functions (LCWFs).
We find that the hard factor of the process can be obtained from single inclusive Drell-Yan production so that one can resum logarithms at high orders in QCD. 
We also discuss the evolution of the GTMDs and the LCWFs. 
\end{abstract}

\begin{keyword}
QCD, Factorization, Drell-Yan, Generalized transverse momentum dependent distributions (GTMDs), Light cone wave functions (LCWFs).  {\it Preprint number:} IPARCOS-UCM-23-023.
\end{keyword}

\end{frontmatter}

\section{Introduction}

The structure of hadrons is parameterized in terms of several multi-dimensional distributions, which encode different correlations between the momenta and spin of the considered partons and their parent hadron. 
Examples of these are the well-known parton distribution functions (PDFs), generalized parton distribution functions (GPDs), or transverse momentum dependent parton distribution functions (TMDs). 
In this work we study the factorization of a process that involves the generalized transverse momentum dependent distributions (GTMDs), that were introduced some time ago \cite{Meissner:2008ay,Meissner:2009ww,Lorce:2013pza,Kanazawa:2014nha}. 

In the literature, several properties of these distributions have been discussed. 
A tree-level analysis shows that GTMDs reduce to TMDs and GPDs~\cite{Meissner:2009ww} after some integration is performed. 
It has been noted that the Fourier transform of the GTMDs is analogous to the classical phase-space distributions, i.e. Wigner distributions~\cite{Ji:2003ak,Belitsky:2003nz,Lorce:2011kd}. Similarly to the case of TMDs, one can consider both quark GTMDs~\cite{Meissner:2008ay,Meissner:2009ww,Lorce:2013pza,Kanazawa:2014nha} and gluon GTMDs~\cite{Martin:1999wb, Martin:2001ms} (which can be interesting for instance for processes with a Higgs particle in the final state \cite{Khoze:2000cy, https://doi.org/10.48550/arxiv.0903.2980}). 

While we have many studies concerning the proof of factorization of cross sections in terms of  GPDs (see e.g. \cite{Muller:1994ses,Radyushkin:1996nd, Ji:1996ek, Goeke:2001tz,Diehl:2003ny,  Ji:2004gf,Belitsky:2005qn, Burkardt:2005hp,Boffi:2007yc,Lorce:2011kd,Guidal:2013rya,Vega:2010ns,Ji:2015qla,Broniowski:2007si, Qiu:2022bpq}) or TMDs (see e.g. \cite{Becher:2010tm,Collins:2011zzd,Echevarria:2011epo,Chiu:2012ir}), we still miss an equivalent proof for the case of GTMDs, even though several proposals have been presented in the last years~\cite{Hatta:2016dxp, Hatta:2016aoc, Hagiwara:2017ofm,Ji:2016jgn,Bhattacharya:2017bvs, Bhattacharya:2018lgm,Boer:2021upt}.
The factorization of a cross-section allows us to identify universal hadronic structures that can be observed in different processes, reduce the model dependence, be able to relate results using the evolution of these distributions, and separate the leading power interesting part of the cross sections from suppressed effects. 

\begin{figure}[h]
\centering
\includegraphics[width=0.3\textwidth]{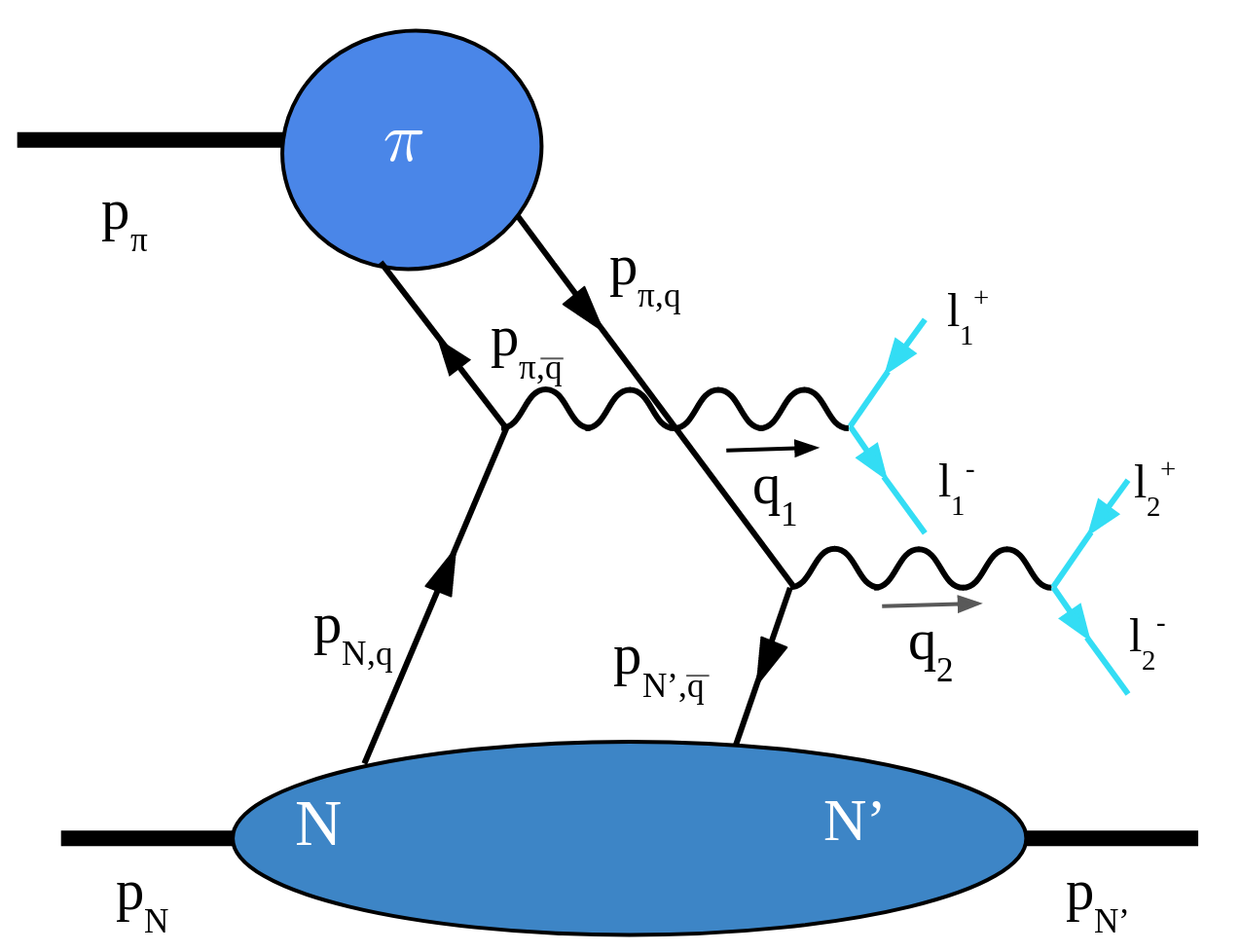}
\label{fig:PiP}
\caption{Exclusive double Drell-Yan at leading power}
\end{figure}

In this work, we show the factorization of an exclusive process proposed for the measurement of GTMDs in \cite{Bhattacharya:2017bvs}, namely 
$\pi N\rightarrow N' \gamma^* \gamma^*\rightarrow N' (\ell^+\ell^-)(\ell^+\ell^-)$, 
where $N,\;N'$ are protons or neutrons (f.i. $\pi^+ p\rightarrow n \gamma^* \gamma^*$, $\pi^- n\rightarrow p \gamma^* \gamma^*$, \dots).  
The process is shown schematically in fig.~\ref{fig:PiP}. 
The interest in this process resides in the fact that it is simple enough to illustrate the basic features of factorization in terms of GTMDs, so that it can work as reference.

The factorization holds in a specific kinematic limit, say
\begin{align}\label{eq:counting}
\frac{|q_{i\perp}|}{\sqrt{q_i^2}}\ll 1
\,,
\end{align}
where $q_i$ with $i=1,2$ are the photons' momenta. 
This ratio allows to establish a power counting in the cross-section, that is consistently taken into account in effective field theories like soft-collinear effective theory SCET~\cite{Bauer:2001yt,Bauer:2002scet,Beneke:2002ph}.   
Using this power counting, eq.~(\ref{eq:counting}), we have only two separated hard interactions, as shown in fig.~\ref{fig:PiP}. 
The contribution due to two photons attached solely to one quark line is power suppressed.
This is because at leading power a hard photon vertex can be coupled only to a collinear and an anti-collinear quark.
We factorize the cross section at leading power in terms of a hard factor, GTMDs and light cone wave functions (LCWFs), and we show that it is consistent with the evolution of these elements. 
The hard factor is purely perturbative, it can be extracted from the analogue hard factor in single inclusive Drell-Yan and it is currently known up to four loops~\cite{Lee_2022, https://doi.org/10.48550/arxiv.2204.02422}. 
This feature, although somehow expected, was never shown explicitly up to now, to our knowledge. 
The evolution of the LCWFs has already been calculated in the literature in  \cite{Ji:2021znw} and we discuss it later in the text.
An important topic of the discussion below is represented by the evolution of GTMDs, calculated in ref.~\cite{Echevarria:2016mrc,Bertone:2022awq} (see also~\cite{Scimemi:2019gge,Vladimirov:2021hdn,Rodini:2022wki} for a general discussion of twist-2 operator anomalous dimensions). 
Both GTMDs and LCWFs have a rapidity scale and an untraviolet scale evolution.
The rapidity evolution (also known as the Collins-Soper kernel) can be deduced too from single inclusive Drell-Yan and it is known at 4 loops in perturbation theory.
\cite{Echevarria:2015byo,Li:2016ctv,Vladimirov:2016dll,Moult:2022xzt, https://doi.org/10.48550/arxiv.2205.02242}.
Using the factorization formulas of this work and the known results from Drell-Yan one can achieve a resummation of large logarithms at a high accuracy.

The paper consists of a derivation of the factorization theorem for the exclusive double DY process in sec.~\ref{sec:Factorization}, followed by a discussion of the QCD evolution of the distributions in sec.~\ref{sec:Evolution}.
A final discussion is then done in the conclusions in sec.~\ref{Conclusions}.

\section{Factorization of the process}
\label{sec:Factorization}

We work in the frame in which the
pion and the nuclei are back to back, so that the hadronic momenta are aligned on the light-cone (collinear or anti-collinear) directions $n$ and $\bar n$, with $n^2=\bar n^2=0$ and $n\cdot \bar n=2$. 
In order to support this kinematics, the dilepton invariant mass $M_{1,2}=q_{1,2}^2$ must be large compared to any other transverse momentum, as specified in eq.~(\ref{eq:counting}). 
In the limit of large $q_{1,2}^2$ and using the SCET power counting the final hadron momentum $p_{N'}$ is collinear to $p_{N}$.
A generic four-vector $v$ is then decomposed as $v^\mu =(v^+, v^-, v_\perp)^\mu
= v^+\frac{n^\mu}{2} + v^-\frac{\bar{n}^\mu}{2} + v_{\perp}$.
The momenta of the hadrons are $p_{N}$ and $p_{N'}$, the momentum of the pion is $p_\pi$ and the momenta of the virtual photons are $q_{1,2}$. 
The momentum conservation of the process is given by 
\begin{align}
p_\pi + p_N &=
p_{N'} + q_1 + q_2
\,.
\end{align}
Two combinations of these momenta will become relevant in the rest of the work, namely:
\begin{align}
\label{eq:kin1}
\Delta &\equiv 
p_{N'}-p_N
\,,\quad
P \equiv
\frac{p_{N'}+p_N}{2}
\,,
\end{align}
where $\Delta$ stands for the {\it momentum transfer} and $P$ stands for the average momentum of the nucleons. 
Alternatively one defines 
$\Delta^\mu = (-2\xi P^+,2\xi P^-,\vec{\Delta}_\perp)$, $\xi \equiv (p_N - p_{N'})^+/(p_N + p_{N'})^+$. 
For definiteness we work in the reference frame with $ p_\pi=( p_\pi^+, p_\pi^-,0)$, $ p_{N}=( p_{N}^+, p_{N}^-,0)$ and $ p_{N'}=( p_{N'}^+, p_{N'}^-, \vec{ p}_{N'\perp})$,  with $ p_\pi^-\gg  p_\pi^+$, $ p_{N,N'}^+\gg |\vec{ p}_{N'\perp}| \gg  p_{N,N'}^- $ and $ p_\pi^- \sim  p_{N,N'}^+$.
If we neglect the masses, as a first approximation we have $ p_\pi\simeq (0, p_\pi^-,0)$ and
$p_{N}\simeq(p_{N}^+,0,0)$.  As a result  of the kinematics of the process, $|t|\equiv|\Delta^2|\sim |P^2|\ll q_{1,2}^2$.

The differential cross-section for the process in the center of mass frame is given by
\begin{equation}
\begin{split}
    \frac{d\sigma}{dM_1^2dy_1d^2\vec{q}_{1\perp}dM_2^2dy_2d^2\vec{q}_{2\perp}} & = d\hat{\sigma}^{\alpha\beta\mu\nu}_{L} W_{\alpha\beta\mu\nu}
\end{split}
\label{eq:Xsec}
\end{equation}
where $d\hat\sigma_L$ collects the leptonic part and overall constants. The 
rapidity of the dilepton couples are given by 
\begin{align}
y_{1,2}=\frac{1}{2}\ln\frac{q_{1,2}^+}{q_{1,2}^-}
\,.
\end{align}
The hadronic tensor $W$, which is the main object of our study, is defined as
\begin{align}
W_{\alpha\beta\mu\nu} &=
\textcolor{black}{\int \frac{d^3 p_{N'}}{2 (2\pi)^3 E_{N'}}}
\int d^4z_{1}d^4z_{2}d^4z_{3}\;  e^{-iq_1\cdot z_1 - iq_2\cdot (z_2-z_3)}
\nn\\
& \times 
\langle\pi N |  \overline{T}\{J^{\dagger}_\alpha(z_1) J^\dagger_\beta(z_2)\} | N'\rangle \,
\langle  N'|T\{J_\mu(z_3) J_\nu(0)\}| \pi N\rangle 
\,,
\end{align}
where $J^\mu = \sum_{q} e_q \bar{\psi}_q \gamma^\mu \psi_q$ is the electromagnetic quark current with $e_q$ as the quark electric charge. 
The extension of this formula for other heavy-color neutral vector bosons can be easily provided, but goes beyond the purposes of this paper, so we do not include it here.

\par
The factorization limit is achieved when one has energetic (hard) photons with momenta scaling as $q_i\sim M_i(1,1,\lambda)$ where $\lambda$ is a small parameter that takes into account that the measured di-lepton transverse momenta are small with respect to photon energies $q_{\perp,i}\ll q_i^\pm$ (see also eq.~(\ref{eq:counting})). 

The leading power factorization can be achieved by applying the effective theory machinery.
Using the SCET power counting, it can be shown that the only contribution to the process is the one depicted in fig.~\ref{fig:PiP}, and all other contributions are power suppressed in $\lambda$.
We do not consider here more sophisticated arguments like in TMD operator expansion \cite{Vladimirov:2021hdn}, where a direct expansion of the QCD Lagrangian is done using the background field method, giving that at leading power they achieve the same final result. 
In all these approaches the original fermion fields are projected onto their collinear dominant modes dressed with collinear radiation, and their position arguments are multipole expanded. 
One has to distinguish the case of fields acting on initial states,
\begin{equation}
\begin{split}
    \psi(z)& \rightarrow  \chi_{n}=W_{n}^{\dagger } (z^+,0,z_\perp) \xi_{n}(z^+,0,z_\perp),\\
      \overline\psi(z) &\rightarrow  \overline{\chi}_{\bar n}= \bar\xi_{\bar n}(z^-,0,z_\perp)W_{\bar n} (z^-,0,z_\perp),\\
     W^\dagger_n(z) &= P\exp\left[ig \int_{-\infty}^0 ds\; n\cdot A_n(z+sn) \right],\\
     W_{\bar n}(z) &= \overline{P}\exp\left[-ig \int_{-\infty}^0 ds\; \bar n\cdot A_{\bar n}(z+s\bar n) \right],\\
\end{split}
\end{equation}
and  the final state
\begin{equation}
\begin{split}
      \overline\psi(z) &\rightarrow  \overline{\chi}_{n}= \bar\xi_{ n}(z^-,0,z_\perp)W_{ n} (z^-,0,z_\perp),\\
     W_{ n}(z) &= P\exp\left[ig \int^{\infty}_0 ds \; n\cdot A_n(z+sn) \right],
\end{split}
\end{equation}
where the subscripts $n,\,\bar n$ indicate collinear and anticollinear fields respectively.
In this way, we match the full QCD electromagnetic currents to the effective current operators. 
For the case where the collinear particle emits (absorbs) a fermion (anti-fermion) and the anti-collinear particle absorbs (emits) an anti-fermion (fermion) we have:
\begin{equation}\label{eq:j}
    J_{\mu, SCET}= \sum_{q} e_{q}  C(Q^2,\mu^2)\bar{\chi}_{q,\bar{n}}S_{\bar{n}}^{\dagger}\gamma_\mu S_{n}\chi_{q,n}
\end{equation}
where $C(Q^2,\mu^2)$ is the hard Wilson coefficient for Drell-Yan. 
This coefficient is well-known from single inclusive Drell-Yan at higher orders in QCD~\cite{Kramer:1986sg,Matsuura:1988sm,Gehrmann:2010ue, Lee_2022, https://doi.org/10.48550/arxiv.2204.02422}.
The soft Wilson lines $S_{n,\bar n}$ appearing in eq.~(\ref{eq:j}), needed to parametrize soft radiation, are all past-pointing in our case and they are defined as
\begin{align}
S_{n}(z) &= P\exp\left[ig \int_{-\infty}^0 ds \; n\cdot A_s(z+s n ) \right]
\,,
\nn\\
S_{\bar n}(z) &= P\exp\left[ig \int_{-\infty}^0 ds \; \bar n\cdot A_s(z+s\bar n ) \right]
\,,
\end{align}
where the subscript $s$ indicates a soft field.
Using Fierz transformations (see~\ref{sec:Appendix} for the Fierz formula of vector currents) we can reshuffle the Dirac indices of the fermion fields in each product of currents, such that the final result is a product of collinear and anti-collinear currents, e.g. 
\begin{equation}
\begin{split}
    \bar{\chi}_{n}^{\prime a}(z_3)\gamma_\mu \chi_{\bar{n}}^{\prime a}(z_3)\; 
    \bar{\chi}_{\bar{n}}^b(0)\gamma_\nu \chi_{n}^b(0) = &\\
     \sum_{\Gamma,\Gamma'} \; C_{\mu\nu}^{\Gamma\Gamma'} \,
    \bar{\chi}_{n}^{\prime a}(z_3)\Gamma \chi_{n}^b(0) \;
    \bar{\chi}_{\bar{n}}^b(0)\Gamma' \chi_{\bar{n}}^{\prime a}(z_3),&
\end{split}
\end{equation}
where $a,\, b$ are color indices and  $\Gamma, \Gamma'=\Gamma_q,\Gamma_{\Delta q},\Gamma_{\delta q}$ with
$\Gamma_q =  \gamma^+$, 
$\Gamma_{\Delta q} =\gamma^+\gamma_5$,
$\Gamma^j_{\delta q} = i\sigma^{j+}\gamma_5$,
and $j=1,2$,
for unpolarized, longitudinally polarized, and transversely polarized quarks \cite{Diehl_2012}. \textcolor{black}{In this equation we also pointed out that the flavor of primed and unprimed fields can be different. Since this feature has a minor role in the factorization that we propose, we omit to differentiate the flavor of fermion fields in the following.
}
Taking into account the scaling of the photons' momenta and the scaling of the fields, one can multipole expand in the leading regions of position space each of the matrix elements. 
We obtain a factorized hadronic tensor with collinear, anti-collinear and soft functions,
\begin{align}
\label{eq:WW}
W_{\alpha\beta\mu\nu} &= H \sum_{\Gamma_1,\Gamma_1',\Gamma_2,\Gamma_2'}
C_{\alpha\beta}^{\Gamma_1\Gamma_1'}
C_{\mu\nu}^{\Gamma_2\Gamma_2'}
 \sum_{\{q\}}  
e_{q_1} e_{q_2} e_{q_3} e_{q_4} W\,.
\end{align}
\begin{figure}[h]
\centering
\includegraphics[width=0.5\textwidth]{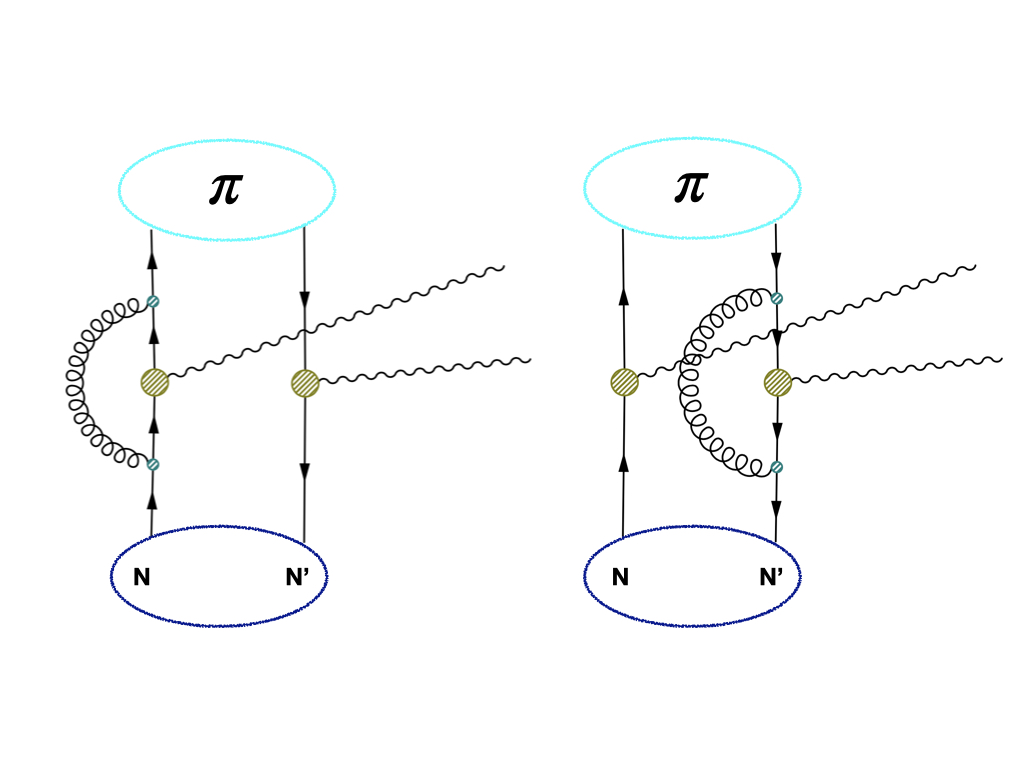}
\vspace{-1cm}
\caption{\label{fig:h1} Diagrams contributing to the hard factor.}
\end{figure}
\begin{figure}[h]
\centering
\includegraphics[width=0.53\textwidth]{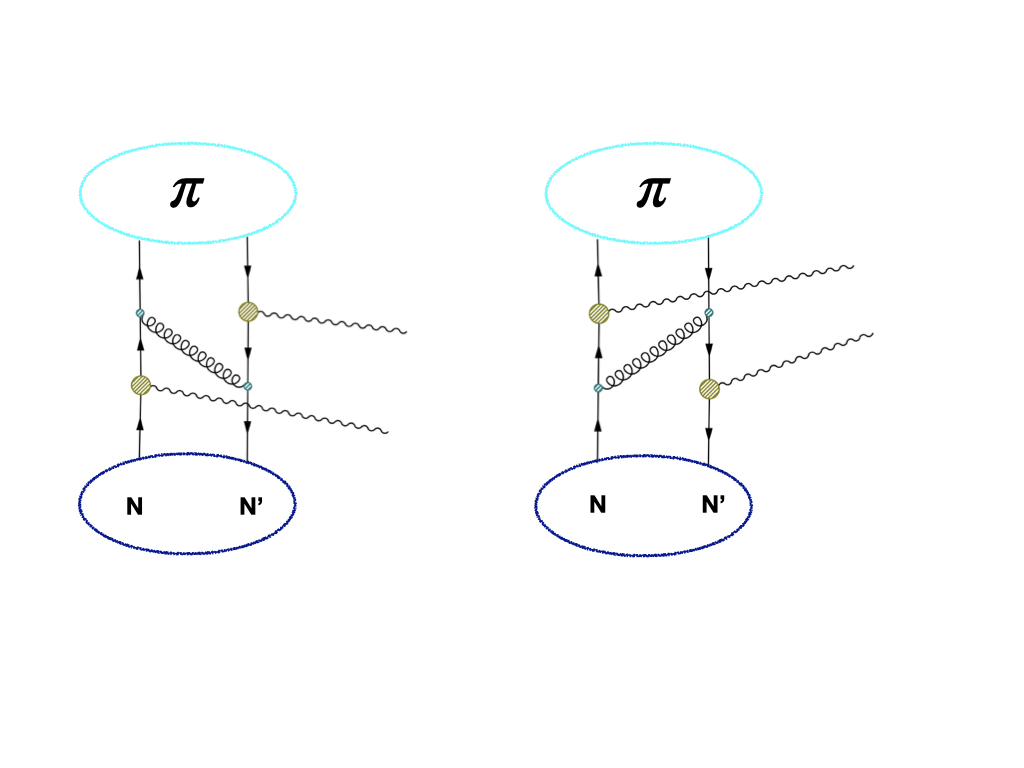}
\vspace{-2cm}
\caption{\label{fig:h2} Diagrams that provide power suppressed contributions to the hard factor.}
\end{figure}
\textcolor{black}{
The diagrams contributing to the hard factor are shown in figs.~(\ref{fig:h1}-\ref{fig:h2}). 
The diagrams in fig.~(\ref{fig:h1}) are well knowm, as they are the same as in single Drell-Yan.
We have performed an explicit calculation of the diagrams in fig.~(\ref{fig:h2}), checking that they are zero at leading power.  
This check has been done performing the integration in coordinate space, following the techniques of \cite{Vladimirov:2021hdn} and considering the incoming and outgoing partons as free particles (which is enough for a leading power computation). 
As a result, we have explicitly checked by a one loop calculation that the factor $H$ satisfies} 
\begin{align}
H = 
|C(Q_1^2,\mu^2)|^2|C(Q_2^2,\mu^2)|^2 
\,,
\label{eq:H}
\end{align}
where the $C(Q^2,\mu^2)$ are the quark current coefficients
\textcolor{black}{ extracted from fig.~(\ref{fig:h1}).}
In DY these coefficients are 
known up to 4-loops \cite{Lee_2022, https://doi.org/10.48550/arxiv.2204.02422}. 
\textcolor{black}{
Our results are compatible with the findings of \cite{Buffing:2017mqm} for double parton scattering, where the same graphs contribute to the hard factor. 
The calculation that we have performed in coordinate space is however original.}

All hadronic matrix elements  in eq.~(\ref{eq:WW}) are contained in the hadronic factor $W$.

Since the soft, collinear and anti-collinear regions do not interact with each other, each matrix element is a color singlet, which means that it is necessary to take the average over color of the collinear and anti-collinear matrix elements. 
Thus we have a contraction of the color indices of the soft Wilson lines,
\begin{align}
W &= \frac{1}{N_c^4}
\int \frac{d^3 p_{N'}}{2 (2\pi)^3 E_{N'}}
\int d^4z_{1}d^4z_{2}d^4z_{3}
e^{-iq_1\cdot z_1 - iq_2 \cdot z_2 + iq_2\cdot z_3}  
\nn\\
& \times 
\langle \pi | 
\overline{\textbf{T}}\{  \bar{\chi}_{\bar{n}}(z_2^+,0^-,\vec{z}_{2,\perp})\Gamma'_{1} \chi_{\bar{n}}(z_1^+,0^-,\vec{z}_{1,\perp})\} |0\rangle
\nn\\
& \times  
\langle 0| \textbf{T}\{
\bar{\chi}_{\bar{n}}(0)\Gamma'_2  \chi_{\bar{n}}(z_3^+,0^-,\vec{z}_{3,\perp})\}
| \pi \rangle
\nn\\
&\times 
\langle 0 | \overline{\textbf{T}} \{  [S_{n}^{\dagger } S_{\bar{n}}](\vec{z}_{1\perp}) [S_{\bar{n}}^{\dagger} S_{n}](\vec{z}_{2,\perp}) \} |0\rangle
\nn\\ \nn
& \times 
\langle 0| \textbf{T}\{ [S_{n}^{\dagger} S_{\bar{n}}](\vec{z}_{3\perp}) [S_{\bar{n}}^{\dagger} S_{n}](0) \} |0 \rangle
\nn\\
&\times \langle N | \overline{\textbf{T}}\{ \bar{\chi}_{n}(0^+,z_1^-,\vec{z}_{1,\perp})\Gamma_{1} \chi_{n}(0^+,z_2^-,\vec{z}_{2,\perp})\} | N' \rangle 
\nn\\
& \times
\langle N'| \textbf{T}\{ \bar{\chi}_{n}(0^+,z_3^-,\vec{z}_{3,\perp})\Gamma_{2} \chi_{n}(0)\}| N \rangle
\,.
\end{align}
Some algebraic passages then follow, as expressing the matrix elements in symmetric positions using the momentum operator,   $J_\mu(z) = e^{iP(z/2)}J_\mu(\textstyle \frac{z}{2}) e^{-iP(z/2)}$
and $e^{-iP\cdot(z^-/2)}|N\rangle = e^{-ip_{N}^+(z^-/2)/2}|N\rangle$), and making the change of variables $z_1 - z_2 \rightarrow  z_1$, $z_3 \rightarrow -z_2$.
With this we obtain
\begin{equation}
\begin{split}
W &= 
\textcolor{black}{\frac{\pi }{N_c^4 E_{N'}}
\delta(q_1^0 + q_2^0+E_{N'}-E_N-E_\pi)}
\int d^4z_{1}d^4z_{2} \\
& \times 
e^{-i(q_1^+ + \Delta^+/2)z_1^-/2}
e^{-i(q_1^- - p_\pi^-/2)z_1^+/2}
e^{+i(\vec{q}_{1\perp} + \vec{\Delta}_{\perp}/2)\cdot \vec{z}_{1\perp}} \\
& \times
e^{-i(q_2^+ + \Delta^+/2)z_2^-/2}
e^{-i(q_2^- - p_\pi^-/2)z_2^+/2}
e^{i(\vec{q}_{2\perp} + \vec{\Delta}_{\perp}/2)\cdot \vec{z}_{2\perp}}
 \\
& \times 
\langle \pi | 
\overline{\textbf{T}}\{  \bar{\chi}_{\bar{n}}(-\textstyle\frac{z_1^+}{2},0^-,-\textstyle\frac{\vec{z}_{1,\perp}}{2})\Gamma'_{1} \chi_{\bar{n}}(\textstyle\frac{z_1^+}{2},0^-,\textstyle\frac{\vec{z}_{1,\perp}}{2})\} |0\rangle\\
& \times  
\langle 0| \textbf{T}\{
\bar{\chi}_{\bar{n}}(\tfrac{z_2^+}{2},0^-,\tfrac{\vec{z}_{2,\perp}}{2})\Gamma'_{2}  \chi_{\bar{n}}(\textstyle -\frac{z_2^+}{2},0^-,-\textstyle\frac{\vec{z}_{2,\perp}}{2})\}
| \pi \rangle\\
&\times \langle 0 | \overline{\textbf{T}} \{  [S_{n}^{\dagger } S_{\bar{n}}](\textstyle\frac{\vec{z}_{1\perp}}{2}) [S_{\bar{n}}^{\dagger} S_{n}](-\textstyle\frac{\vec{z}_{1,\perp}}{2}) \} |0\rangle\\
& \times \langle 0| \textbf{T}\{ [S_{n}^{\dagger} S_{\bar{n}}](-\tfrac{\vec{z}_{2\perp}}{2}) [S_{\bar{n}}^{\dagger} S_{n}](\tfrac{\vec{z}_{2\perp}}{2}) \} |0 \rangle\\
&\times \langle N | \overline{\textbf{T}}\{ \bar{\chi}_{n}(0^+,\textstyle\frac{z_1^-}{2},\textstyle\frac{\vec{z}_{1,\perp}}{2})\Gamma_{1} \chi_{n}(0^+,-\textstyle\frac{z_1^-}{2},-\textstyle\frac{\vec{z}_{1,\perp}}{2})\} | N' \rangle \\
& \times
\langle N'| \textbf{T}\{ \bar{\chi}_{n}(0^+,-\tfrac{z_2^-}{2},-\tfrac{\vec{z}_{2,\perp}}{2})\Gamma_2 \chi_{n}(0^+,\tfrac{z_2^-}{2},\tfrac{\vec{z}_{2,\perp}}{2})\}| N \rangle .
\label{eq:Wbracket}
\end{split}
\end{equation}
Having $W$ so factorized is still an intermediate step. In fact the matrix elements that compose it have overlapping contributions that should be eliminated with a zero-bin subtraction, similarly to the TMD case.  Presently we have two soft functions, which makes this case different from the TMD factorization. The zero-bin subtraction is regularization dependent and in our discussion we will refer to the modified $\delta$-regulator~\cite{Echevarria:2016scs} for practical examples. 

\subsection{Kinematic variables and unsubtracted distributions}

It is useful to introduce a set of variables that distinguish the mean value of momenta and their deviation from average variables and also to point out approximate values for them. Similar definitions exist also in ~\cite{Bhattacharya:2017bvs} so we limit to list them here,
\begin{align}
\label{eq:kinvar}
&q_1^++q_2^+=
-\Delta^+ + p_{\pi}^+ \simeq 
-\Delta^+ \equiv 2\xi P^+ 
\nn\\
&q_1^+-q_2^+= 2x_P P^+ \nn \\
&q_1^-+q_2^-=
-\Delta^-+p_\pi^-\simeq 
p_\pi^- 
\nn\\    
&q_1^++\frac{\Delta^+}{2} \simeq
x_P P^+
\nn\\
&q_2^++\frac{\Delta^+}{2} \simeq
-x_P P^+
\nn\\
&q_2^- \equiv
x_\pi p_\pi^-
\nn\\
&q_1^-\simeq
(1-x_\pi) p_\pi^-
\,.
\end{align}
In these equations the symbol "$\simeq$" is used when hadronic masses are neglected, which we always assume in the present derivation. For the kinematic of the process  we have a quark with collinear  momentum $P^+ (x_P+\xi)\geq 0$ and an anti-quark with $P^+(x_P-\xi)\leq 0$, that is
\begin{align}
\label{eq:erbl}
 -\xi\leq x_P\leq\xi .
\end{align}
This interval is the so-called ERBL region \cite{Lepage:1980fj,Efremov:1978rn}.

The kinematic variables  just introduced together with integration over the longitudinal positions allow us to write the hadronic factor $W$ in 
eq.~(\ref{eq:Wbracket}) as 
\begin{align}
W &= \nn
\textcolor{black}{\frac{\pi }{4N_c^4 E_{N'}}
\delta(q_1^0 + q_2^0+E_{N'}-E_N-E_\pi)}
\\ \nn
& \times \int d^2z_{2\perp}\, e^{-i\frac{1}{2}\Delta\vec q_\perp
\cdot \vec{z}_{2\perp}}
\\ \nn
& \times 
\phi^{[\Gamma'_2]}(x_\pi, z_{2\perp})\,
S(z_{2\perp})\,
w^{[\Gamma_2]}(x_P, z_{2\perp}; \xi,\vec \Delta_\perp)
\\ \nn
& \times \int d^2z_{1\perp}\, e^{i\frac{1}{2}\Delta\vec q_\perp
\cdot \vec{z}_{1\perp}}
\\ 
&\times
\phi^{*[\Gamma'_1]}(x_\pi,z_{1\perp})\,
S(z_{1\perp})\,
w^{*[\Gamma_1]}(x_P,z_{1\perp}; \xi,\vec \Delta_\perp)
\,,
\end{align}
where $\Delta_\perp = -(q_{1\perp}+q_{2\perp})$ 
and 
$\Delta q_\perp \equiv q_{1\perp}-q_{2\perp}$.
The hadronic distributions that appear above, i.e. the unsubtracted GTMDs and LCWFs, and the soft functions, are defined as
\begin{align}
&w^{[\Gamma_2]}(x_{P},z _{2\perp};\xi, \vec\Delta_\perp) =
\int \frac{dz^-_2}{4\pi}
e^{i\frac{1}{2} x_P P^+ z^-_2} 
\nn\\ 
& 
\hspace{1cm}
\times
\langle N'|\textbf{T}\{  \bar{\chi}_{n}(0^+,-\tfrac{z^-_2}{2} ,-\tfrac{\vec{z}_{2\perp}}{2})   \Gamma_2
\chi_{n}(0^+,\tfrac{z^-_2}{2},\tfrac{\vec{z}_{2\perp}}{2})\}| N\rangle
\,,
\nn\\
& \phi^{[\Gamma'_2]}_{\pi}(x_\pi,\vec{z}_{2\perp}) =
\int \frac{dz_2^+}{4\pi}
e^{-i\frac{1}{2}(x_\pi-\frac{1}{2})p_\pi^- z_2^+} 
\nn\\ 
& 
\hspace{1cm} 
\times
\langle 0| \textbf{T}\{
\bar{\chi}_{\bar{n}}(\tfrac{z^+_2}{2}, 0^-,\tfrac{\vec{z}_{2\perp}}{2})\Gamma'_2 \chi_{\bar{n}}(-\tfrac{z_2^+}{2}, 0^-, -\tfrac{\vec{z}_{2\perp}}{2})
\}| \pi \rangle
\,,
\nn\\
&S(\vec{z}_{\perp})=
\langle 0|\overline{\textbf{T}} \{[S_{n}^{\dagger } S_{\bar{n}}](0^+,0^-,-\tfrac{\vec{z}_{\perp}}{2}) [S_{\bar{n}}^{\dagger} S_{n}](0^+,0^-,\tfrac{ \vec{z}_{\perp}}{2}) \} |0\rangle
\,.
\end{align}

\subsection{Definition of GTMDs and LCWFs}

In the definitions of all these functions, we have omitted the regulators necessary for their computation. 
In practice, they all contain rapidity divergences that need to be treated.
The presence of the rapidity regulators in the collinear and anti-collinear functions can be removed considering the zero-bin that accounts for the overlap of (anti-)collinear and soft modes.
Notice that the interactions between collinear and anti-collinear modes are power suppressed so that the zero-bin subtraction works very similarly to the single inclusive Drell-Yan case. For instance, using the $\delta$-regulator the zero-bin and the soft function coincide, and these are formally the same as in single inclusive Drell-Yan. 
Like in the case of TMDs, as a final result, one obtains a dependence of GTMDs and LCWFs on the rapidity evolution scale $\zeta$ and the renormalization scale $\mu$, associated with the regularization of ultraviolet divergences.

Schematically, to obtain an expression without rapidity divergences one needs to split the product of soft factors as
\begin{align}
\label{eq:soft_factors_split}
\raggedleft 
&S(\vec{z}_{1\perp};\mu^2,\delta^+\delta^-)\, 
S(\vec{z}_{2\perp};\mu^2,\delta^+\delta^-) = 
\nn
\\
&\quad\quad
=
\textstyle \sqrt{S\left(\vec{z}_{1\perp};\mu^2,(\delta^{+})^2
\displaystyle
\frac{\zeta_1}{ p_{N,q}^+  p^+_{N^\prime,{\bar q}}}
\right)}
\nn
\\
&\quad\quad
\times
\textstyle \sqrt{S\left(\vec{z}_{1\perp};\mu^2,(\delta^-)^2 
\displaystyle
\frac{\bar \zeta_1}{p_{\pi, q}^-p_{\pi, \bar q}^-} \right)}
\nn
\\
&\quad\quad
\times 
\sqrt{S\left(\vec{z}_{2\perp};\mu^2,(\delta^{+})^2
\displaystyle
\frac{\zeta_2}{ p_{N,q}^+  p^+_{N^\prime,\bar q}}\right)}
\nn\\ 
&\quad\quad
\times 
\textstyle \sqrt{S\left(\vec{z}_{2\perp};\mu^2,(\delta^-)^2 
\displaystyle \frac{\bar \zeta_2}{ p_{\pi,q}^-p_{\pi,\bar q}^-}
\right)}
\,,
\end{align}
where we explicitly use the partonic momenta of quarks and anti-quarks inside hadrons, $p_{i}$, consistently with their appearance in the perturbative calculations of the unsubtracted GTMDs and LCWFs. 
In eq.~(\ref{eq:soft_factors_split}) we have considered the product of soft functions and not them individually, as this is the quantity that enters in the cross-section. 
From the last equation we obtain that the product of soft factors is factorized with the condition
\begin{align}
\label{eq:zetai}
\zeta_1\bar\zeta_1 
\zeta_2\bar\zeta_2
&=
(p_{\pi, q}^-p_{\pi, \bar q}^-  p_{N,q}^+ p^+_{N^\prime,\bar q})^2
\,,
\end{align}
given that the soft function is linear in the rapidity logarithms to all orders in perturbation theory (see e.g. \cite{Echevarria:2012js}).
The resulting well-defined GTMDs, free from rapidity divergences, entering the factorized cross-section are
\begin{align}
     W_{N'N}^{[\Gamma_2]}(x_P,\vec{z}_{2\perp};\xi,\vec \Delta_\perp; \mu^2, \zeta_2) &= 
     \nn 
     \\ 
     &
     \hspace{-1cm}
     = w_{N'N}^{[\Gamma_2]}(x_P,\vec{z}_{2\perp}; \xi,\vec \Delta_\perp;\mu^2, \delta^+) \nn 
     \\
     &
     \hspace{-1cm}
     \times
     \sqrt{S\left(\vec{z}_{2\perp};\mu^2,(\delta^{+})^2 \displaystyle\frac{\zeta_2}{p_{N,q}^+ p^+_{N^\prime,\bar q}} \right)}\nn \,,
     \\
     W_{NN'}^{*[\Gamma_1]} (x_P,\vec{z}_{1\perp} ;\xi,\vec \Delta_\perp;\mu^2 ,\zeta_1)&= 
     \nn 
     \\
     &
     \hspace{-1cm} 
     = w_{NN'}^{*[\Gamma_1]}(x_P,\vec{z}_{1\perp};\xi,\vec \Delta_\perp;\mu^2,\delta^+)
     \nn
     \\
     &
     \hspace{-1cm} 
     \times
     \sqrt{S\left(\vec{z}_{1\perp};\mu^2,(\delta^{+})^2 \displaystyle
     \frac{\zeta_1}{p_{N,q}^+ p^+_{N^\prime,\bar q}} \right)}\,.
\end{align}

Analogously, for the LCWFs we have
\begin{align}
    \Phi_{\pi}^{[\Gamma'_2]} (x_\pi,\vec{z}_{2\perp};\mu^2,\bar\zeta_2) & =
    \nn
    \\ 
    &\hspace{-1cm}
    \phi_{\pi}^{[\Gamma'_2]}(x_\pi,\vec{z}_{2\perp};\mu^2,\delta^-)
    \nn 
    \\
    &\hspace{-1cm}
    \times
    \sqrt{S\left(\vec{z}_{2\perp};\mu^2,(\delta^-)^2 \displaystyle
    \frac{\bar \zeta_2}{p_{\pi,q}^-p_{\pi,\bar q}^-}
    \right)} \nn \,,
    \\
    \Phi_{\pi}^{*[\Gamma'_1]}(x_\pi,\vec{z}_{1\perp};\mu^2,\bar\zeta_1 ) 
    &=
    \nn
    \\
    &
    \hspace{-1cm}
    = \phi_{\pi}^{*[\Gamma'_1]}(x_\pi,\vec{z}_{1\perp} ;\mu^2,\delta^-)
    \nn 
    \\ 
    &
    \hspace{-1cm} 
    \times
    \sqrt{S\left(\vec{z}_{1\perp};\mu^2,(\delta^-)^2 \displaystyle 
    \frac{\bar \zeta_1}{p_{\pi,q}^-p_{\pi,\bar q}^-} \right)}
    \,.
\end{align}
In the equations above the distributions  $w_{NN'}$ and $\phi_\pi$ are understood to be zero-bin subtracted.
Using the rapidity divergence-free definitions we obtain
\begin{align}
W &= \nn
\textcolor{black}{\frac{\pi }{4N_c^4 E_{N'}}
\delta(q_1^0 + q_2^0+E_{N'}-E_N-E_\pi)}
\\ \nn
& \times \int d^2z_{2\perp} e^{-i\frac{1}{2}\Delta\vec q_\perp
\cdot \vec{z}_{2\perp}}
\\ \nn
& \times 
\Phi^{[\Gamma'_2]}(x_\pi,\vec  z_{2\perp};\mu^2,\bar\zeta_2)\,
W^{[\Gamma_2]}_{N'N}(x_P, \vec z_{2\perp}; \xi,\vec \Delta_\perp;\mu^2,\zeta_2)
\\ \nn
& \times \int d^2z_{1\perp} e^{i\frac{1}{2}\Delta\vec q_\perp
\cdot \vec{z}_{1\perp}}
\\ 
&\times
\Phi^{*[\Gamma'_1]}(x_\pi,\vec z_{1\perp};\mu^2,\bar\zeta_1)\,
W^{*[\Gamma_1]}_{NN'}(x_P,\vec z_{1\perp}; \xi,\vec \Delta_\perp;\mu^2,\zeta_1)
\,,
\label{eq:Wb}
\end{align}
which shows explicitly the scale dependence of each factor.

For completeness, we also write the factorized hadronic tensor as a convolution in momentum space:
\begin{align}
W &=
\frac{\pi }{4N_c^4 E_{N'}}
\delta(q_1^0 + q_2^0+E_{N'}-E_N-E_\pi)
\nn\\
& \times 
\int d^2\vec k_{ n 2\perp}
d^2\vec k_{\bar n 2\perp}
\delta^{(2)}
(\textstyle \frac{\Delta\vec q_\perp}{2} + \vec k_{\bar n 2\perp} - \vec k_{n 2\perp} )
\nn\\
&
\times 
\Phi^{[\Gamma'_2]}(x_\pi,  \vec k_{\bar n 2\perp} ;\mu^2,\bar\zeta_2)\; 
\nn
W^{[\Gamma_2]}_{N'N}(x_P, \vec k_{n 2\perp}; \xi,\vec \Delta_\perp;\mu^2,\zeta_2)
\nn\\
& \times 
\int d^2\vec k_{ n 1\perp}
d^2\vec k_{\bar n 1\perp}
\delta^{(2)}
(\textstyle \frac{\Delta\vec q_\perp}{2} + \vec k_{\bar n 1\perp} - \vec k_{n 1\perp} )
\nn\\ 
&\times
\Phi^{*[\Gamma'_1]}(x_\pi,\vec k_{\bar n 1\perp};\mu^2,\bar\zeta_1)\;
W^{*[\Gamma_1]}_{NN'}(x_P, \vec k_{ n 1\perp}; \xi,\vec \Delta_\perp;\mu^2,\zeta_1)
\,,
\label{eq:Wk}
\end{align}
where we have used
\begin{align} \nn
    &W^{[\Gamma_2]}_{N'N}(x_P, \vec k_{n 2\perp}; \xi,\vec \Delta_\perp;\mu^2,\zeta_2) = \\
    \nn
    &
    \hspace{1cm}
    =
    \frac{1}{2} \int \frac{dz_2^- d^2\vec{z}_{2\perp}}{(2\pi)^3}
    e^{i x_P P^+z_2^-/2  -  i\vec k_{n2\perp}\cdot \vec{z}_{2\perp}} \\
    \nn
    & 
    \hspace{1cm} 
    \times 
    \langle N'|\textbf{T}\{  \bar{\chi}_{n}(0^+,-\tfrac{z^-_2}{2} ,-\tfrac{\vec{z}_{2\perp}}{2})   \Gamma_2
    \chi_{n}(0^+,\tfrac{z^-_2}{2},\tfrac{\vec{z}_{2\perp}}{2})\}| N\rangle
    \,, \\ 
    & \Phi^{[\Gamma'_2]}(x_\pi,  \vec k_{\bar n 2\perp} ;\mu^2,\bar\zeta_2) =
    \nn
    \\ 
    & 
    \hspace{1cm} 
    =
    \frac{1}{2} \int  \frac{dz_2^+ d^2\vec{z}_{2\perp}}{(2\pi)^3}
    e^{-i(x_\pi - \tfrac{1}{2})p_\pi^-z_2^+/2 + i \vec{k}_{\bar{n}2\perp} \cdot \vec{z}_{2\perp} }
    \\
    \nn
    &
    \hspace{1cm} 
    \times 
    \langle 0| \textbf{T}\{
    \bar{\chi}_{\bar{n}}(\tfrac{z^+_2}{2}, 0^-,\tfrac{\vec{z}_{2\perp}}{2})\Gamma'_2 \chi_{\bar{n}}(-\tfrac{z_2^+}{2}, 0^-, -\tfrac{\vec{z}_{2\perp}}{2})
    \}| \pi \rangle .
\end{align}
The final form of the factorized hadronic tensor, and thus the cross-section, is provided by eqs.~(\ref{eq:Wb}-\ref{eq:Wk}).

\section{QCD evolution of GTMDs and LCWFs}
\label{sec:Evolution}

The consistency of the factorization requires that the sum of the anomalous dimensions of all the factors in the cross-section cancels. 
It is important to provide this check also in our case. 

The anomalous dimensions have been explicitly calculated in the literature. 
In fact, we have that the anomalous dimension of the hard factor $H=|C(Q_1^2,\mu^2)|^2|C(Q_2^2,\mu^2)|^2$ can be obtained from
\begin{align}
\frac{d}{d\ln\mu} C(Q^2,\mu^2)= \Gamma_{\rm cusp}\ln\frac{Q^2}{\mu^2}+\gamma_V
\,,
\end{align}
where $\Gamma_{\text{cusp}}$ and $\gamma^V$ are cusp and non-cusp anomalous dimensions known up to third~\cite{Moch:1999eb,Moch:2005id,Moch:2005tm} and fourth order in $\alpha_s$~\cite{Lee_2022, https://doi.org/10.48550/arxiv.2204.02422}.
Then for the GTMDs and the LCWFs we find respectively
\begin{align}
&\frac{d}{d\ln\mu} 
\ln W_{N'N}^{[\Gamma]}(x_P, z_{\perp};\xi, \vec\Delta_\perp;\mu,\zeta  ) = 
\gamma_W \left( a_s,\mu,\zeta\right)
\,,
\nn\\ 
& 
\frac{d}{d\ln\mu}\ln \Phi_{\pi}^{[\Gamma]}(x,\vec{z}_{\perp};\mu, \zeta) =  \gamma_\Phi \left( a_s,\mu, \zeta\right)
\,.
\label{eq:m}
\end{align}
Furthermore, the evolution equation in the rapidity parameter of the GTMDs and LCWFs is given by the rapidity anomalous dimension 
$D(\vec{z}_{\perp};\mu)$ of Drell-Yan~\cite{Collins:2011zzd, GarciaEchevarria:2011rb,Chiu:2012ir,Becher:2010tm}:
\begin{align}
& \frac{d}{d\ln \zeta}\ln W_{N'N}^{[\Gamma]}(x_P,\vec{z}_{\perp}; \xi,\vec \Delta_\perp;\mu,\zeta )
= -D(\vec{z}_{\perp};\mu)
\,, 
\nn\\ 
& 
\frac{d}{d\ln \zeta}
\ln 
\Phi_{\pi}^{[\Gamma]}(x,\vec{z}_{\perp};\mu,\zeta) = 
-D(\vec{z}_{\perp};\mu) 
\,.
\label{eq:z}
\end{align}
The calculation of the anomalous dimension for each distribution can be found in \cite{Echevarria:2016mrc,Bertone:2022awq,Ji:2021znw} and yields
\begin{align}
\label{eq:adwp}
\gamma_W \left( a_s,\mu,\zeta\right) = 
\gamma_\Phi \left( a_s,\mu,\zeta\right) = \gamma_{\Phi^*} \left( a_s,\mu,\zeta\right) =
\gamma \left( a_s,\mu,\zeta\right)
\,, 
\end{align}
with $a_s=\alpha_s/(4\pi)$ and
\begin{align}
\gamma \left( a_s,\mu,\zeta\right) =
\Gamma_{\rm cusp}\ln\frac{\mu^2}{\zeta}-\gamma_V
\,.
\end{align}
Combining these equations with the one of the hard coefficient $H$, one obtains the condition for the cancellation of the sum of all anomalous dimensions:
\begin{align}
\zeta_1\bar\zeta_1 \zeta_2\bar\zeta_2&= Q^4_{1} Q^4_2
\,,
\end{align}
which is consistent with $p_{N,q}^+p_{N',\bar q}^+=q^+_1q_2^+$, $p_{\pi,q}^-p_{\pi,\bar q}^-=q^-_1q_2^-$
and eq.~(\ref{eq:zetai}) and $Q_i^2=q_i^+q_i^-$ (with i=1,2)
\footnote{
The anomalous dimensions are a property of operators, so they can be deduced from twist-2 computations  \textcolor{black}{that we explicitly checked} in
\cite{Vladimirov:2021hdn,Rodini:2022wki} (see also \cite{Belitsky:2005qn,Balitsky:2022vnb}). At one loop one finds
\begin{align}
\label{eq:ads}\nn
  \gamma_W(a_s,\mu^2,\zeta)&=
     \gamma(a_s,\mu^2,\zeta)+
     2 a_s 
    \displaystyle \ln\frac{(q^+)^2}{k_{N,q}^+k_{N',\bar q}^+},\\
\nn
 \gamma_\Phi(a_s,\mu^2,\zeta)&= \gamma(a_s,\mu^2,\zeta)+2a_s\ln\frac{(q^-)^2}{k_{\pi,q}^-k_{\pi,\bar q}^-},
 \nn
 \end{align}
where $k_i^\pm$ are the momenta associated with quark operators and $q^\pm$ are high-energy scales that appear in the generic soft matrix element splitting 
$$ S(\delta^+\delta^-,z_\perp^2)=R(\delta^+/q^+,\zeta,z_\perp^2)
R(\delta^-/q^-,\zeta,z_\perp^2).
$$
The extra logarithms appearing in these anomalous dimensions
are canceled when $(q^+)^2=k_{N,q}^+k_{N',\bar q}^+$ and $(q^-)^2=k_{\pi,q}^-k_{\pi,\bar q}^-$ so that the final result agrees with our eq.~(\ref{eq:adwp}). 
Notice that with this choice of high-energy scales $q^\pm$, all the anomalous dimensions are real and thus valid for both the ERBL and DGLAP regions.
In our settings, this is consistent with
$(q^+)^2=q_1^+q_2^+=p_{N,q}^+p_{N',\bar q}^+=k_{N,q}^+k_{N',\bar q}^+$ and 
$(q^-)^2=q_1^-q_2^-=p_{\pi,q}^-p_{\pi,\bar q}^-=k_{\pi,q}^-k_{\pi,\bar q}^-$.
}.

Using eq.~(\ref{eq:m}-\ref{eq:z}) one can write the GTMDs with single inclusive DY evolution kernel:
\begin{equation}
\begin{split} 
     & W_{N'N}^{[\Gamma]}(x_{P},\vec{z}_{\perp};\xi,\vec \Delta_\perp;\mu,\zeta ) = R(\vec{z}_{\perp}; \mu,\zeta,\mu_0,\zeta_{0}) \\ & \times
    W_{N'N}^{[\Gamma]}(x_{P},\vec{z}_{\perp};\xi,\vec \Delta_\perp;\mu_0,\zeta_{0}  ),
\end{split}
\end{equation}
with
\begin{equation}
\begin{split}
    R(\vec{z}_{\perp}; \mu,\zeta,\mu_0,\zeta_{0}) &= \left(\frac{\zeta}{\zeta_{0}}\right)^{-D(\vec{z}_{\perp};\mu_0)}\\
    & \times \exp{\left[\int_{\mu_0}^{\mu}\frac{d\bar{\mu}}{\bar{\mu}}\ \gamma \left( \alpha_s(\bar \mu),\ln\frac{\bar \mu^2 }{\zeta }\right) \right]}
    \,,
\end{split}
\end{equation}
and likewise for the LCWFs.
We end this section by emphasizing that the evolution of both GTMDs and LCWFs is spin independent, and thus the same for all leading-twist distributions.
In the case of gluon distributions, one would need to use the analogous anomalous dimension and rapidity evolution kernel for gluons (see e.g. \cite{Echevarria:2015uaa}), and again they would be spin independent.

\section{Conclusion}
\label{Conclusions}

In this work we have obtained a factorized formula for the process   $\pi N\rightarrow N' \gamma^* \gamma^*\rightarrow N' (\ell^+\ell^-)(\ell^+\ell^-)$ that was proposed in \cite{Bhattacharya:2017bvs}  to show how GTMDs could be extracted from a physical process.
The factorization has been achieved using soft-collinear effective theory, and it has been derived under the kinematic conditions of eq.~(\ref{eq:counting}).
This derivation is somewhat similar to the factorization of double parton scattering~\cite{Diehl:2017wew}, since there are two hard interactions between the colliding hadrons. 
The main difference concerns the color structure of the final matrix elements (GTMDs and LCWFs) which are just color singlets in our case. 
This difference is important and crucial for the achievement of the final result.

The factorization has consistently included the evolution of all functions.  
The extraction of the hard part of the process allows us to use evolution equations up to higher orders in perturbation theory, which is one of the main results of the paper.

A final comment concerns the relation between GTMDs and GPDs~\cite{Meissner:2009ww,Lorce:2013pza}. 
It is often stated that the integration over the transverse variable of GTMDs allows one to recover the corresponding GPDs. 
This statement {\it per se} is not correct.
In this sense, the situation is very similar to the relation among TMDs and PDFs~\cite{Collins:1981va,Collins:2016hqq,Ebert:2022cku}, where an approximate enforcing of this statement is studied.
We also leave for the future a more detailed study of the relation between GTMDs and TMDs (see~\cite{Gurjar:2021dyv}), as well as the study of other processes which factorize in terms of GTMDs.

\section*{Acknowledgements}
We thank A. Vladimirov for useful discussions on the subject of this work.
This project is supported by the Spanish Ministry grant PID2019-106080GB-C21. 
This project has received funding from the European Union Horizon 2020 research and innovation program under grant agreement Num. 824093 (STRONG-2020).
P.A.G.G. is supported by the FPI contract PRE2020-094385.

\appendix
\section{Appendix}
\label{sec:Appendix}
We provide here the Fierzing transformation necessary for the product of electromagnetic currents.
\begin{align}
4(\gamma^\mu)_{ij}(\gamma^\nu)_{kl}&=\nonumber
\left[ {\bf 1}_{il}{\bf 1}_{kj}
+(i\gamma_5)_{il}(i\gamma_5)_{kj}-(\gamma^\alpha)_{il}(\gamma_\alpha)_{kj}
\nn
\right. \\ 
&\left. -(\gamma^\alpha\gamma_5)_{il}(\gamma_\alpha\gamma_5)_{kj}
+\frac{1}{2} (i\sigma^{\alpha\beta}\gamma_5)_{il}(i\sigma_{\alpha\beta}
\gamma_5)_{kj}\right]g^{\mu\nu} \nonumber\\ \nn
&+(\gamma^{\{ \mu})_{il}(\gamma^{\nu\} })_{kj}+
(\gamma^{\{ \mu}\gamma_5)_{il}(\gamma^{\nu\} }\gamma_5)_{kj}
\nn \\ & \nn
-\frac{1}{2} 
(i\sigma^{\alpha\{\mu}\gamma_5)_{il}(i{\sigma^{\nu\}} }_\alpha \gamma_5)_{kj}\\ & \nn
-\frac{i \epsilon^{\mu\nu\lambda\eta}}{2}[(i\sigma^{\lambda\eta}\gamma_5)_{il}{\bf 1}_{kj}-
{\bf 1}_{il}(i\sigma^{\lambda\eta}\gamma_5)_{kj}]\\ &\nn
+(i\sigma^{\mu\nu}\gamma_5)_{il}{\gamma_5}_{kj}+{\gamma_5}_{il}(i\sigma^{\mu\nu}\gamma_5)_{kj}\\ &
+i \epsilon^{\mu\nu\alpha\beta}\Big[(\gamma^\alpha \gamma_5)_{il}(\gamma^\beta)_{kj}+(\gamma^\alpha)_{il}(\gamma^\beta \gamma_5)_{kj} \Big]
\end{align}

\bibliographystyle{utphys}
\biboptions{sort&compress}
\bibliography{GTMD}

\end{document}